\documentstyle[aps,twocolumn,epsf]{revtex}
\input epsf

\newcommand{\be}{\begin{eqnarray}}
\newcommand{\ee}{\end{eqnarray}}

\newcommand{\slL}{\raise.15ex\hbox{$/$}\kern-.53em\hbox{$L$}}
\newcommand{\slP}{\raise.15ex\hbox{$/$}\kern-.53em\hbox{$P$}}
\newcommand{\slR}{\raise.15ex\hbox{$/$}\kern-.53em\hbox{$R$}}
\newcommand{\slQ}{\raise.15ex\hbox{$/$}\kern-.53em\hbox{$Q$}}
\newcommand{\slK}{\raise.15ex\hbox{$/$}\kern-.53em\hbox{$K$}}
\newcommand{\slSigma}{\raise.15ex\hbox{$/$}\kern-.53em\hbox{$\Sigma$}}
\newcommand{\slcalP}{\raise.15ex\hbox{$/$}\kern-.63em\hbox{$\cal P$}}
\begin{document}

\title{\bf{Soft dilepton production and hard thermal loops \thanks{Based on 
work done in collaboration with P.~Aurenche, F.~Gelis and R.~Kobes. 
Talk presented at the
      TFT'98 conference, Regensburg, Germany, 10-14 august 1998. }}}
\author{H.~Zaraket}
\address{Laboratoire de Physique Th\'eorique LAPTH,\\
BP110, F-74941, Annecy le Vieux Cedex, France}
\date{\today}

\maketitle

\begin{abstract}

The Hard Thermal Loop expansion, is an attractive theory, but it reveals 
difficulties when one uses it as a perturbative scheme. To illustrate 
this we use the HTL expansion to calculate the two loop corrections for soft 
virtual photons. The resulting corrections are of the same order as the 
1-loop correction. 
\end{abstract} 

\vskip 2mm
LAPTH-98/702, hep-ph/9810246

\section{Introduction}
The production of soft photons (energy $\sim gT$, where $g\ll 1$ is the strong coupling
constant) in a hot quark-gluon plasma at equilibrium has 
been  studied by many groups, using thermal field theory  
\cite{collin1,ALTH89,Pisar90,CLEDAD,AurenGKP1}, 
or a semi-classical approach as in 
\cite{CleymGR1,CleymGR2,GolovR1}. In this talk I will study the case of  soft 
virtual photon (with invariant mass $Q^{2} \sim g^{2}T^{2}$) at two-loop 
level using the Hard Thermal Loop (HTL) expansion.
The two loop diagrams are found to be of the same order of magnitude as the one 
loop result. This unexpected result solves the problem of the absence of
bremsstrahlung in the HTL expansion, while bremsstrahlung seems to 
contribute dominantly when using the semi-classical approach
 \cite{CleymGR1,CleymGR2,GolovR1}.

\section{Motivations and problems}
\label{sect:motive}
\begin{figure}[ht]
  \centerline{
    \mbox{\epsfysize=2cm\epsffile{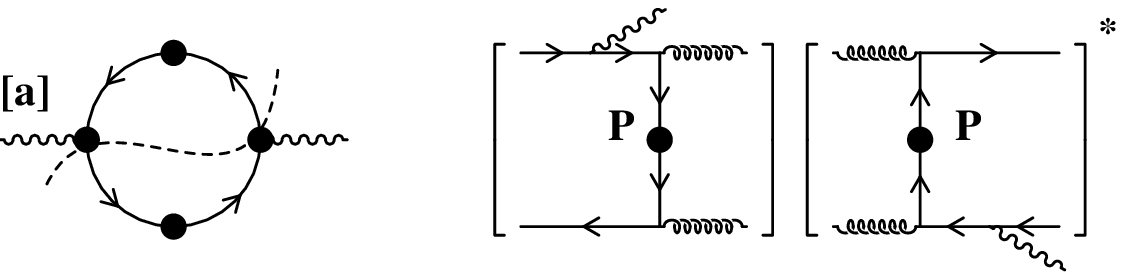}}
    }
  \vskip 2mm
  \centerline{
    \mbox{\epsfysize=2cm\epsffile{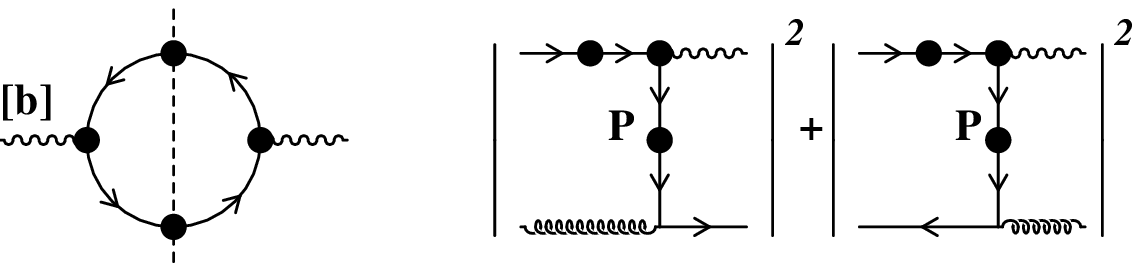}}
    }
  \caption{\footnotesize{1-loop contribution to the soft photon production and
  some of the different cuts responsible for different physical processes; in 
  [b] one of the two fermion momenta is space like while the other is time like.
  }}
  \label{fig:1loop}
\end{figure}

\begin{figure}[htbp]
  \centerline{
    \mbox{\epsfysize=2cm\epsffile{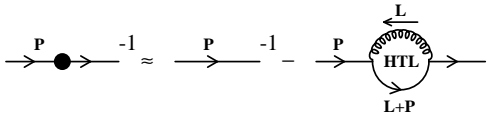}}}
  \caption{\footnotesize{The inverse of the fermion effective propagator.}}
  \label{fig:self}
\end{figure}

\begin{figure}[htbp]
  \centerline{
    \mbox{\epsfysize=2cm\epsffile{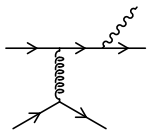}}}
  \caption{\footnotesize{Bremsstrahlung photon production.}}
  \label{fig:brem}
\end{figure}

The starting point will be the bare thermal field theory. The imaginary part of the 
photon polarization which is proportional to the invariant dilepton 
production rate is found to be:
\begin{eqnarray}
{\rm Im}\Pi^\mu_\mu &\sim & e^{2}Q^{2}(1-2n_{F}(\frac{Q}{2}))\nonumber\\
                    &\sim & e^{2}g^{3}T^{2}\qquad{\rm for}\quad q,q_{o}\sim gT\;
		    .
\label{eq:bare}
\end{eqnarray}
The additional power of $g$ in ${\rm Im}\Pi^\mu_\mu$  arises from the fact 
that the soft photon is produced by annihilation of a soft 
quark-antiquark pair (each of momentum $Q/2$). Other processes such as 
$q\rightarrow\gamma q$ are not allowed since the quark is massless. 
Having a soft momentum scale, the HTL expansion \cite{BraatP1,BraatP2,FrenkT1,FrenkT2} 
 demands the use of effective vertices and propagators. The soft virtual
photon production was one of the first applications of the HTL resummation scheme
\cite{Pisar90}: the authors of \cite{Pisar90} have calculated the imaginary part of the one
loop diagram of Fig.~\ref{fig:1loop}. 
In Fig.~\ref{fig:1loop} we  have depicted some of the 
cuts{\footnote{The thermal cutting rules \cite{KobesS1,KobesS2} is a 
method that one can use to evaluate
the imaginary part of a Green function.}} that 
give relevant physical amplitudes.
Cuts [a] and [b] of the one loop diagram in Fig.~\ref{fig:1loop} can be
approximated by :
\begin{equation}
 {\rm Im}\,\Pi^\mu{}_\mu(Q)_{|{\bf [a,b]}} \sim e^2 g^3
 {{T^2}}\int\limits_{{gT}}^{{T}}\frac{dp}{p}\; ,
\label{eq:1loop}
\end{equation}
where p is the quark momentum in the loop, where we have used explicitly 
$q_o\sim gT$.
A few remarks are worth saying about this result:\\
$\bullet$ The HTL is built to handle soft momentum problems; in the hard
momentum limit, it tells that we recover the bare theory which is {\it a priori}
what one needs in {\it most} cases. For example if we take a fermion with momentum 
$P$, the inverse of its effective propagator is given by the sum of the  
inverse bare propagator ($\slP$) and the one-loop correction as shown in
Fig.~\ref{fig:self}: to evaluate
this correction one should apply the HTL approximation. Thus in the one-loop
correction in Fig.~\ref{fig:self} we use $L+P \approx L$, which is justified if
$P$ is soft and the loop momentum $L$ is hard, but when $P$ becomes hard this
approximation is no more correct, then the HTL estimate of the one-loop
correction of the effective propagator is incomplete in the hard momentum limit.   
$\bullet$ The processes shown in Fig.~\ref{fig:1loop} come from one loop 
corrections to the bare vertices and propagators. On the other hand the integral 
$\int\limits_{gT}^{T}\frac{dp}{p}\sim\ln(1/g)$ in Eq.~\ref{eq:1loop}, indicates 
that the quark momentum runs from a soft scale ($gT$) to a hard scale ($T$), 
{\it i.e.} we need the hard limit of the one loop correction which is 
incomplete in the HTL approximation as we mentioned before.\\
$\bullet$ The importance of processes coming from cut [a] in Fig~\ref{fig:1loop}
 can be understood from the result of Cleymans et al  
\cite{CleymGR1,CleymGR2,GolovR1}, where they calculated bremsstrahlung photon
production (Fig.~\ref{fig:brem}) using a semi classical approach: their result
yields ${\rm Im}\,\Pi^\mu{}_\mu(Q)\sim e^2 g^3 T^2$ which is of 
the same order as Eq.~\ref{eq:bare} and Eq.~\ref{eq:1loop}.
$\bullet$ The bremsstrahlung photon production (Fig.~\ref{fig:brem}) is mediated
by the gluon exchange, while that of cut [a] of Fig.~\ref{fig:1loop} is mediated
by the exchange of quark, and it is reasonable to think that a complete
calculation should include both processes. The
natural question that arises is: how can one reconcile bremsstrahlung with the 
HTL scheme?

\section{2-loop diagrams}
\begin{figure}[htbp]
  \centerline{
    \mbox{\epsfysize=2cm\epsffile{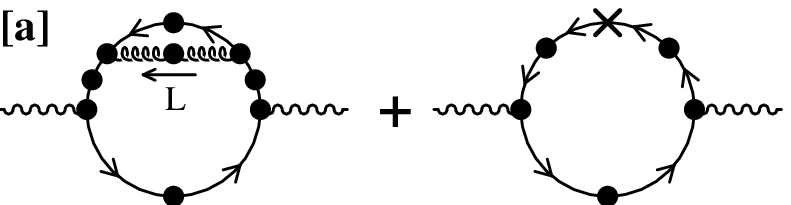}}
    }
  \vskip 1mm
  \centerline{
    \mbox{\epsfysize=2cm\epsffile{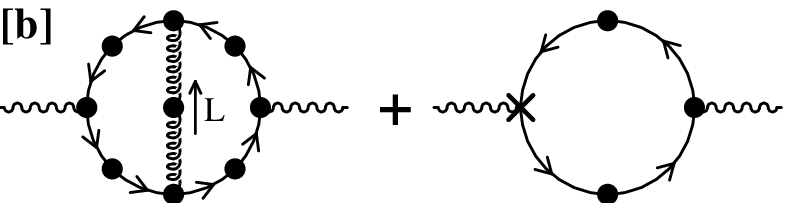}}
    }
  \caption{\footnotesize{Two-loop contributions involving 
      bremsstrahlung processes. A black dot denotes an effective
      propagator or vertex. Crosses are HTL counterterms.}}
  \label{fig:2loops}
\end{figure}

\begin{figure}[htbp]
  \centerline{
    \mbox{\epsfysize=2cm\epsffile{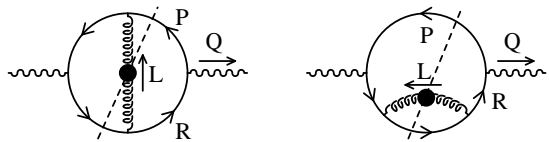}}}
  \caption{\footnotesize{Simplified two-loop contributions involving 
      bremsstrahlung processes.}}
  \label{fig:2loopsimple}
\end{figure}

The solution to the problems mentioned in Sect.~\ref{sect:motive} comes when one 
looks at 2-loop diagrams in the HTL expansion. In two loop diagrams we have more
powers of $g$, coming from the $qqg$ vertices, then to compensate additional powers
of $g$ which may come from the phase space we have to have a hard phase space. 
In the catalogue of 2-loop diagrams we
skip those which contain vertices with no bare analogue: these vertices turn out to
be negligible when one of the vertex lines becomes hard. 
The remaining diagrams are shown in Fig.~\ref{fig:2loops}. As discussed in F.
Gelis's talk \cite{gelistak}, calculating two loop diagrams in the HTL
expansion , one should pay attention to the counterterms in order to
avoid double-counting (see also \cite{Aurngkz98} for more details.).     
We explore the case where all the quark momenta are hard. Assuming hard quark momentum we can replace 
the effective vertices by the bare ones, and take the bare propagator for the 
quark. The simplified version of Fig.~\ref{fig:2loops} is represented in 
Fig.~\ref{fig:2loopsimple}. In the same figure we have depicted the relevant 
cuts that enable one to calculate  the corresponding contribution in the framework 
of the thermal cutting rules. 
In what follows we will take the photon to be static {\it i.e.}
$Q^{\mu}=(q_{o},{\bbox 0})$. For virtual photon the production rate per unit 
time and per unit volume is related to the imaginary part of the retarded polarization 
tensor of the photon as \cite{Weldo3,GaleK1}:
\begin{equation}
  {{dN}\over{dtd{\bbox x}}}=-
  {{dq_od{\bbox q}}\over{12\pi^4}}\;
  {\alpha\over{Q^2}}\,n_{_{B}}(q_o)\,
  {\rm Im}\,\Pi^{^{R}}{}_\mu{}^\mu(q_o,{\bbox q})\; .
  \label{virtualphot}
\end{equation}
The gluon propagator appears linearly in ${\rm Im}\,\Pi_\mu{}^\mu$, then we can 
write:
\begin{equation}
{\rm Im}\,\Pi_\mu{}^\mu\sim {\rm Im}\,\Pi_\mu{}^\mu\left |_{{L^{2}<0}} + 
{\rm Im}\,\Pi_\mu{}^\mu\left |_{{L^{2}>0}}\right .\right .\; ,
\label{eq:pmL2}
\end{equation}
where $L$ is the momentum flow through the gluon.\\
Such separation is motivated by the fact that regions $L^{2}<0$ and $L^{2}>0$ 
contain different physical processes. While $L^{2}<0$ gives bremsstrahlung of
Fig.~\ref{fig:brem}, which is a possible candidate to reconcile bremsstrahlung
with the HTL expansion, 
$L^{2}>0$ gives annihilation and Compton scattering. Bremsstrahlung is a new process, then we can guarantee 
that it will give a positive contribution that should be added to the one-loop
result.
Moreover we can apply different approximations in each of
these regions. We can treat each region separately and at the end we add the two
contributions. 
In the next sections we will discuss in detail the region $L^{2}<0$, while  we will give
some indicative results for $L^{2}>0$.

\section{Bremsstrahlung}
The matrix element ${\cal M}$ of diagrams [a] and [b] of
Fig.~\ref{fig:2loopsimple} , can be written as (see \cite{Aurngkz98} for the 
complete expression):
\begin{equation}
{\cal M}\sim {\cal M}_{1}L^{2} + {\cal M}_{2}\; .
\label{eq:matrix}
\end{equation}
Writing ${\cal M}$ in this form exhibits  some of the terms that are absent at
one loop level in the HTL expansion. Indeed the gluon appearing in the
effective propagator and vertices in the one loop
diagram are taken to be light-like ($L^{2}=0$) by construction of the HTL 
resummation scheme, then ${\cal M}_{1}$ will be a pure 2-loop contribution. 

\subsection{Extraction of logarithmic behavior}
\label{sect:logext}
We can obtain analytically the leading logarithmic
behavior in $\ln(1/g)$ of ${\rm Im}\,\Pi_\mu{}^\mu$ at 2-loop order.
By momentum power counting we can isolate the term which gives the logarithm: it
happens that this term comes from ${\cal M}_{1}$, enforcing the previous expectation
that bremsstrahlung contribution is a pure two loop contribution.\\
The contribution of this leading term can be written schematically as:
\begin{equation}
{\rm Im}\,\Pi^{^{R}}{}_\mu{}^\mu(q_o,{\bbox 0})\approx {\rm A}\times
e^2g^2\left(\frac{1}{q_o}\right)m^2_{\rm
g}\int\limits_0^{T}dp\int\limits_{f(m_{\rm g},q_o)}^{T}\frac{dl}{l}\; .
\end{equation} 
{\bf Comments:}\\
$\bullet$ A is a constant that depends on the color factors, and it has no $g$
dependence.\\
$\bullet$ $e^2$ comes from the $qq\gamma$ couplings.\\
$\bullet$ $g^2$ comes from the $qqg$ couplings.\\
$\bullet$ $\frac{1}{q_o}$ comes from the uncut quark-propagator denominators, 
for example the momentum $R$ in Fig.~\ref{fig:2loopsimple} can be approximated
by: $R^2=2q_o p$.\\
$\bullet$ The soft gluon thermal mass $m_{\rm g}^2\equiv
g^2T^2[N+N_{_{F}}/2]/9$, comes from the gluon spectral density.\\
$\bullet$ The ultraviolet cutoff $T$ is provided by the statistical weights.\\
$\bullet$ The integral $\int\limits_0^{T}dp$ indicates the hard $p$ behavior.\\
$\bullet$ $f(m_{\rm g},q_o)$ is a function of dimension one. This function
depends on both $q_o$ which appears as a kinematical infrared cut-off in the
integral over $dl$ and $m^2_{\rm g}$ which appears in the gluon spectral density
and can also play the role of an infrared cut-off for the same integral.\\
$\bullet$ The integral $\int\limits_{f(m_{\rm g},q_o)}^{T}\frac{dl}{l}$
points out that $l$ vary between a soft scale ($f\sim gT$) and a hard scale ($T$).\\
The cutoff $f$ can be simplified by taking $q_o\ll m_{\rm g}\sim gT\ll T$, then
the larger cutoff $m_{\rm g}$ will be the relevant cutoff. After this
simplification we get: 
\begin{eqnarray}
  {\rm Im}\,\Pi^{^{R}}{}_\mu{}^\mu(q_o,{\bbox 0})\approx
  -{{3NC_{_{F}}e^2g^2}\over{8\pi^3}} {{m^2_{\rm g} T}\over {q_o}}
  \ln\left({{T^2}\over{m^2_{\rm g}}}\right)\; .
\end{eqnarray}

\begin{figure}[t]
  \centerline{\mbox{\epsfysize=3cm\epsffile{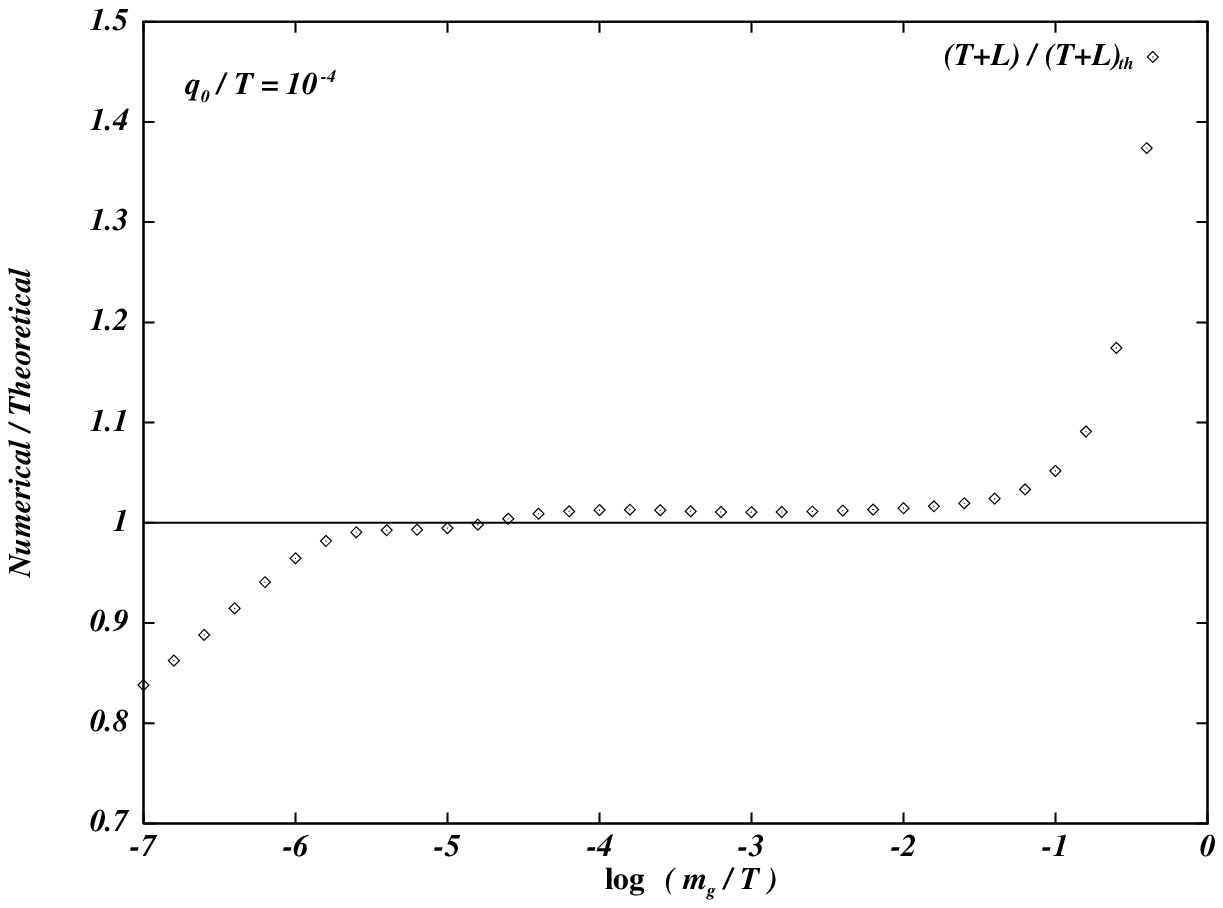}}\hglue 1mm
    \mbox{\epsfysize=3cm\epsffile{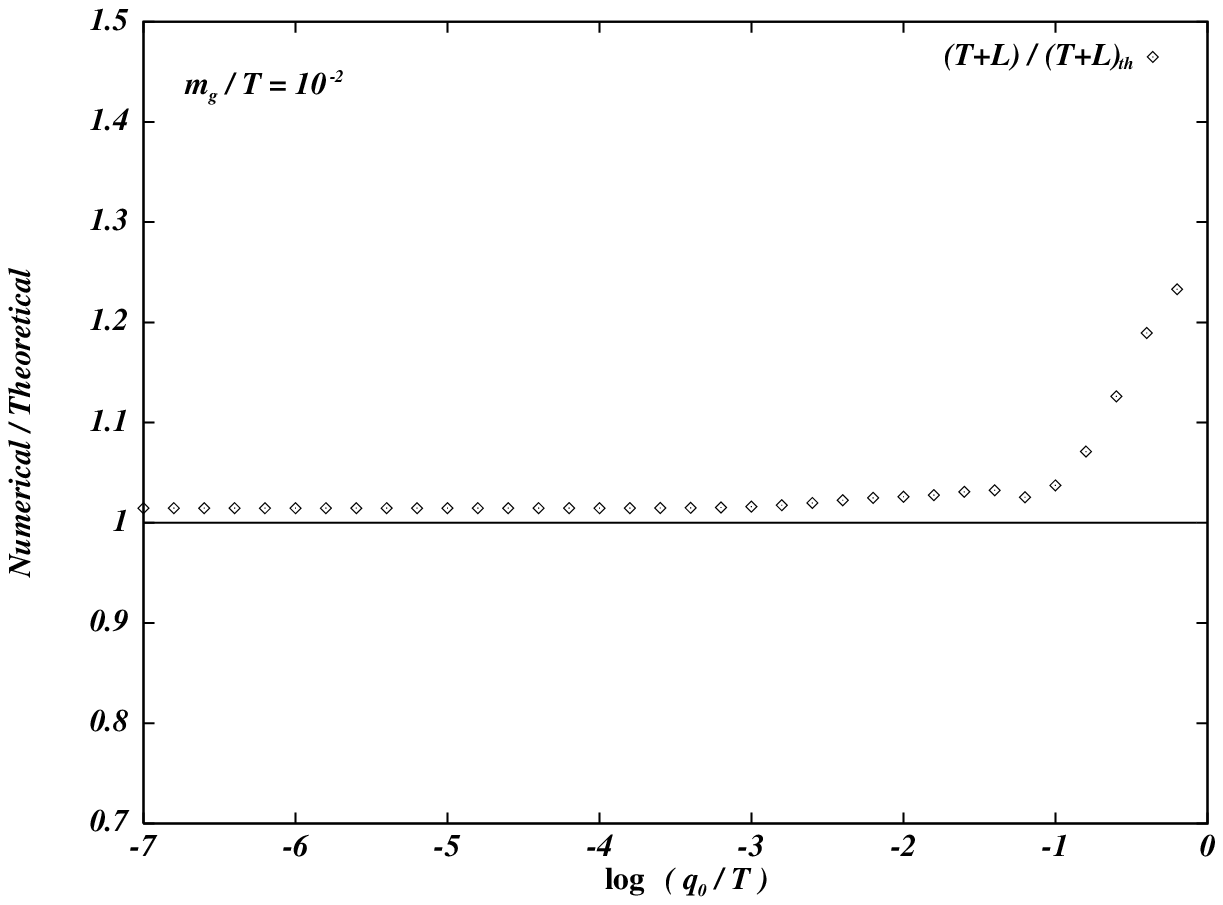}}}
  \caption{\footnotesize{Comparison of numerical estimates of the complete
      matrix element with the simple theoretical expression obtained in
      Eq.~(\ref{eq:staticfinal}). Both plots show the ratio
      ``Numerical/Theoretical''. On the left plot, $q_o/T$ is fixed at
      $10^{-4}$ and we look at the variations with $m_{\rm g}/T$ ({\it i.e.~}
      with $g$). On the right plot, $m_{\rm g}/T$ is fixed at $10^{-2}$ and
      the photon energy varies from ultra-soft energies to hard ones.}}
  \label{fig:numeric}
\end{figure}

The production rate is then given by:
\begin{eqnarray}
  \left.{{dN}\over{dtd{\bbox x}}}\right|_{\rm bremss}\approx
  {{dq_od{\bbox q}}\over{8\pi^6}}NC_{_{F}}
  \alpha^2\Big(\sum_{f}e^2_f\Big)
  \left({{m_{\rm g}\over {q_o}}}\right)^2\nonumber\\
  \times\left({{gT}\over {q_o}}\right)^2 \ln\left({{T^2}
      \over{m^2_{\rm g}}}\right)\; ,
  \label{eq:staticfinal}
\end{eqnarray}
where the sum runs over the flavor of the quarks in the loop ($e_f$ is
the electric charge of the quark of flavor $f$, in units of the electron
electric charge).\\
Numerical estimates of the unapproximated expression of 
${\rm Im}\,\Pi_\mu{}^\mu$  verify that the condition $q_o\ll m_{\rm g}$ is only
a {\it technical} condition as we get a reasonable accuracy even for 
$q_o/m_{\rm g}\sim 10$. On the other hand  $m_{\rm g}/T$ must be smaller than 0.1 in
order to have an agreement between the approximate and the complete expression
with an accuracy better than 5 \% (see Fig.~\ref{fig:numeric}).
\subsection{Comparison with other approaches}
\subsubsection{Braaten et al. results}
The production rate of soft static photons has been
evaluated by Braaten, Pisarski and Yuan in \cite{Pisar90} at one-loop in the HTL
scheme. To compare
Eq.~\ref{eq:staticfinal} with the result  of \cite{Pisar90} in the domain 
$q_o\ll m_{\rm g}\sim gT\ll T$ for which our expression is justified, one has
to isolate from their result the processes which have the same nature and 
spectrum (in $\frac{1}{q_o^{4}}$) as the bremsstrahlung, {\it i.e} those of 
Fig.~\ref{fig:1loop}-[a]. Applying the same approximation we did in
Sect.~\ref{sect:logext}, we can easily extract the $\ln(1/g)$ in BPY's result:
\begin{equation}
  \left.{{dN}\over{dtd{\bbox x}}}\right|_{\rm 1-loop}\!\!\!\!\!\approx
  {{dq_od{\bbox q}}\over{12\pi^4}}N\alpha^2\Big(\sum_{f}e^2_f\Big)
  \left({{m_{_{F}}\over {q_o}}}\right)^4 \ln\left({{T^2}
      \over{m^2_{_{F}}}}\right)\; ,
\label{eq:bpy}
\end{equation}
where $m^2_{_{F}}= g^2 C_{_{F}} T^2/8$ is the soft quark thermal mass.\\
{\bf Comments:}
\begin{itemize}
\item[$\bullet$] The production rate in Eq.~\ref{eq:bpy} and 
Eq.~\ref{eq:staticfinal} have the same spectrum in $\frac{1}{q_o^{4}}$.
\item[$\bullet$] $m^2_{_{F}}$ plays the role of an infrared regulator in 
Eq.~\ref{eq:bpy}, while $m_{\rm g}$ is the regulator in 
Eq.~\ref{eq:staticfinal}. 
This reflects the different nature of both results, an exchanged fermion in the 
first, and an exchanged gluon in the second.
\item[$\bullet$] Comparing the two results we obtain the ratio:
\begin{equation}
  {{\left.{{dN}}\right|_{\rm bremss}}
    \over{\left.{{dN}}\right|_{\rm 1-loop}}}\approx
  {{32}\over{3\pi^2}}{{N+N_{_{F}}/2}\over{C_{_{F}}}}\; ,
\end{equation}
which for 2 light flavors and 3 colors becomes
\begin{equation}
  {{\left.{{dN}}\right|_{\rm bremss}}
    \over{\left.{{dN}}\right|_{\rm 1-loop}}}\approx
  {{32}\over{\pi^2}}\sim 3.2\; .
\end{equation}
This ratio is rather large, which means that bremsstrahlung is definitely
an essential contribution to the soft static photon production rate by a hot
plasma.
\end{itemize}

\subsubsection{Cleymans et al. results}
We return to the semi-classical treatment used in
\cite{CleymGR1,CleymGR2,GolovR1}. In their approach they took into account
the effect of multiple scattering of the quark in the plasma. To compare the two 
loop result with their result one need to ``undo" the effect of rescattering 
and consider only one collision of the quark in the plasma. We apply the same
simplification as they did {\it i.e.}:
\begin{itemize}
\item[$\bullet$] The scattering of the particles is treated as in vacuum (the
dynamics), while the plasma effects are introduced for the kinematics only.
\item[$\bullet$] The energy of the quarks or gluons is much larger than the 
temperature so that Boltzman distributions are used for particles entering the 
interaction region and a factor 1 is assigned to those leaving it. As a  
by-product one can neglect the photon momentum compared to the momenta of the 
constituents in the plasma.
\item[$\bullet$] To screen the forward singularity they introduced a 
phenomenological Debye mass $m_{_D}$, which we will take to be $m_{\rm g}$
when we compare their result with Eq.~\ref{eq:staticfinal}.
\end{itemize}
Applying the above approximations we get for the production rate of the lepton
pair at rest the following expression:
\begin{equation}
  \displaystyle{{{dN}\over{dtd{\bbox x}}}} \approx 
  {d q_o d{\bbox q}\over{3\pi^6 }}{\alpha^2\alpha_s^2}{d_{f} d}
  \Big(\sum_{f}e^2_f\Big){T^4\over q_o^4} \ln\left({{T^2}
      \over{m^2_{_{D}}}}\right)\; ,
\label{eq:semic}
\end{equation}
where $d_f=2_s\times 3_c = 6$ and 
$d=\frac{4}{9}\ 2\times 2_f\times 2_s\times 3_c + 2_s\times 8_c = 26+{2\over 3}$ 
are the degeneneracy factors introduced in \cite{CleymGR2}.\\
From Eq.~\ref{eq:semic} we see that the semi-classical approach give the same 
spectrum in $\frac{1}{q_o^{4}}$ as expected.\\ 
Comparing with Eq.~\ref{eq:staticfinal} we find for two flavors:
\begin{equation}
  {\left.{dN}\right|_{\rm semi-class} \over \left.{dN}\right|_{\rm bremss}}
  \approx {15 \over \pi^2}\; ,
\end{equation}
{\em i.e.~}the semi-classical over-estimates the rate
of production by about $50 \%$. This difference appears to come from the
various approximations they used {\footnote{The Boltzman approximation is a 
favorite candidate of such an overestimate.}}.

\section{Preliminary results for time like gluon}
In this section we give partial results for the $L^{2}>0$
case \cite{prog} and the arguments that we will give are based on the leading 
logarithmic behavior only.\\ 
The basic physical processes for $L^{2}>0$ are Compton and annihilation, then
naively we expect double counting with the one loop diagram, in the
region where the imaginary part of the one loop diagram gives similar processes 
(Fig.~\ref{fig:1loop}-[b]). To cure this problem of double counting , we have to 
calculate the counterterms depicted on the right of  Fig.~\ref{fig:2loops}.
Indeed, when the gluon in the loop becomes hard, we have a hard loop that
reproduce what is already included in the one-loop diagram via the effective
vertices and propagators .\\ 
The structure of ${\rm Im}\,\Pi_\mu{}^\mu$ allows to decompose it as the
sum of two pieces, one part reflecting the hard gluon momentum behavior 
(${\rm Im}\,\Pi_{1}$) where we can apply the hard $L$ approximation, and the 
other reflecting the hard fermion behavior (${\rm Im}\,\Pi_{2}$):
\begin{equation}
{\rm Im}\,\Pi\simeq {\rm Im}\,\Pi_{1} + {\rm Im}\,\Pi_{2}\; .
\end{equation}

\subsection{Hard $L$ region}
By isolating hard $L$ terms we apply indirectly the HTL approximation, then one
may expect a fair compensation between ${\rm Im}\,\Pi_{1}$ and the counterterms.
${\rm Im}\,\Pi_{1}$ reveals soft $p$ sensitivity, which means that it is no more
allowed to use the bare propagator for the quark, thus we have to use the
effective one. At leading logarithm order we can take a constant mass for the quark
which will be the asymptotic quark thermal mass which is twice the soft quark
thermal mass $m_{_{F}}$. This mass will play along with $q_0$ the role of an 
infrared regulator. \\
Then the contribution of ${\rm Im}\,\Pi_{1}$ to the production rate is
approximated by:
\begin{equation}
 \left. \frac{dN}{d^4xdq_o d^3{\bbox q}}\right|^{{\rm Im}_{1}}\approx\frac{N\sum_{f}
  e_{f}^2}{12\pi^4}\frac{\alpha^2 m_{f}^2}{q_o^2}\ln\left(\frac{T}
  {g(m_{_{F}},q_o)}\right)\; ,
\label{eq:logim1prod}
\end{equation}
where $g$ is a function of $m_{f}$ and $q_o$. 
To evaluate the imaginary part of the counter term diagrams (CT) we follow 
\cite{Pisar90}, where the authors relate the imaginary part of the effective
vertex to the spectral density function. Then the 
imaginary part of the counter term diagrams may be 
expressed in terms of the spectral density of the quark propagator. After a
cumbersome calculation, the logarithmic behavior of the counter terms can be 
estimated by:
\begin{equation}
 \left. \frac{dN}{d^4xdq_o d^3{\bbox q}}\right|^{\rm CT}\approx -\frac{N\sum_{f}
  e_{f}^2}{12\pi^4}\frac{\alpha^2 m_{_{F}}^2}{q_o^2}\ln\left(\frac{2Tq_o}
 {h(m_{_{F}},q_o)}\right)\; ,
\label{eq:logcount}
\end{equation}
where $h$ is a function of $m_{{F}}$ and $q_o$.\\
The only difference between Eq.~\ref{eq:logim1prod} and Eq.~\ref{eq:logcount} is
that they have different infrared regulator, adding both Eqs. we get a logarithm
of argument of ${\cal O}(1)$, which remains finite, even in the limit of 
vanishing thermal masses {\footnote{A limit to be used later on.}}, then we can
neglect the total contributions coming from ${\rm Im}\,\Pi_{1}$ and the
counterterms compared to $\ln(1/g)$ that we will get from ${\rm Im}_{2}$ .

\subsection{Hard $p$ region}
In the hard $p$ limit we are lead to the same type of analysis as for
$L^{2}<0$ where there is no need for the effective propagator, thus we use the bare 
propagator for the quark. The different behavior of the spectral 
density for $L^2<0$ and
$L^2>0$ leads to different behavior of the terms in the matrix element, for
example the term that gives the leading logarithm in ${\rm Im}\;\Pi_{2}$ comes 
from ${\cal M}_{2}$ in Eq.~\ref{eq:matrix}. Different behavior will lead to 
different spectrum in each case.\\
Taking a constant thermal mass for the gluon, we can obtain analytically the 
leading logarithmic behavior of ${\rm Im}_{2}$:
\begin{equation}
 {\rm Im}\;\Pi_{2}(q_o,{\bbox 0})\approx {{NC_{_{F}}e^2g^2}\over{4\pi^3}}{q_o}T 
 \ln\left(\frac{T}{m_{\rm g}
 \left(\frac{m_{\rm g}}{q_o}+\frac{q_o}{m_{\rm g}}\right)}\right)\; .
\label{eq:im1log}
\end{equation}
The production rate is then given by:
\begin{equation}
   \left. \frac{dN}{d^4xdq_o d^3{\bbox q}}\right|^{{\rm hard\;\;}
   p}\!\!\approx\!\!
   \frac{2N\sum e_{f}^2}{3\pi^6}\frac{\alpha^2m_{_{F}}^2}{q_o^2}\ln
   \left(\frac{T}{m_{\rm g}\left(\frac{m_{\rm g}}{q_o}+\frac{q_o}
   {m_{\rm g}}\right)}\right)\; ,
\label{eq:pl2loop}
\end{equation} 
which has a $\frac{1}{q_o^2}$ spectrum compared to the $\frac{1}{q_o^4}$ of
bremsstrahlung.
\subsection{Comparison with Braaten et al. result}
To compare Eq.~\ref{eq:pl2loop} with the one loop result obtained in 
\cite{Pisar90}, we consider only the kinematical region which gives rise to 
processes shown in Fig.~\ref{fig:1loop}-[b],
{\it i.e.} the region where one of the quarks in the loop is space like while
the other is time like. 
Applying the same approximation done to obtain Eq.~\ref{eq:im1log} we get for
the diagram [b] of Fig.~\ref{fig:1loop}:
\begin{equation}
   \left. \frac{dN}{d^4xdq_o d^3{\bbox q}}\right|^{\rm 1loop}\approx
   \frac{N\sum e_{f}^2}{12\pi^4}\frac{\alpha^2m_{_{F}}^2}{q_o^2}\ln
   \left(\frac{2Tq_o}{{\rm Max}(q_o^2,m_{_{F}}^2)}\right)\; ,
\label{eq:oneloop}
\end{equation}
which has the same spectrum in $\frac{1}{q_o^{2}}$ as the production rate at 
two loops in  Eq.~\ref{eq:pl2loop}. 
Adding Eqs.~\ref{eq:pl2loop} and \ref{eq:oneloop} we get for the annhilation and
Compton scattering up to two loops:
\begin{eqnarray}
  \frac{dN}{d^4xdq_o d^3{\bbox q}}\approx\frac{N\sum e_{f}^2}{3\pi^6}
  \frac{\alpha^2m_{f}^2}{q_o^2}\left\{\frac{\pi^2}{4}\ln\left(\frac{2Tq_o}
  {{\rm Max}(q_o^2,m_{F}^2)}\right)\right.\nonumber\\
  \left.\!\!\!\!\!\!\!\!\!\!\!\!\!+2\ln\left(\frac{T}{m_{\rm g}\left(
  \frac{m_{\rm g}}{q_o}+\frac{q_o}{m_{\rm g}}\right)}\right)\right\}\; .
\label{eq:logtot}
\end{eqnarray}
The ratio of the two-loop contribution to the one-loop result is given by:
\begin{equation}
   {{\left.{{dN}}\right|_{\rm 2-loops}}
    \over{\left.{{dN}}\right|_{\rm 1-loop}}}\approx
    {{8}\over{\pi^2}}\sim 0.81\; .
\label{eq:compton}
\end{equation}
This signifies that the two-loop contribution is essential.
Being calculated with $p$ taken hard from the beginning, ${\rm Im}\,\Pi_{2}$
will complete the contribution coming from the 1-loop result which assumes $p$ to
be soft compared to the momentum of the hard loop in the resumed propagators and
vertices. This cures (at least at leading logarithmic behavior) the problem 
mentioned in Sect.~\ref{sect:motive}, that the hard thermal loop does not give 
the complete estimate of the loop corrections to the
effective quantities in the limit of hard momentum. To sum-up, the one-loop
diagram takes care of the soft $p$ region, while the two-loop diagrams with 
the counter terms complements the one loop estimate by taking care of the hard 
$p$ region.   

To verify the above result we take the thermal masses to zero in the 
logarithms in Eq.~\ref{eq:logtot}:
\begin{equation}
  \frac{dN}{d^4xdq_o d^3{\bbox q}}\approx\frac{N\sum e_{f}^2}{3\pi^6}
  \frac{\alpha^2g^2C_{_{F}}}{8q_o^2}(\frac{\pi^2}{4}+2)\ln(T/q_o)\; .
\end{equation}
This coincides with the production rate calculated in \cite{ALTH89} at two-loop
level using the bare thermal field theory.

\section{Further problems}
\begin{figure}[htbp]
  \centerline{
    \mbox{\epsfysize=2cm\epsffile{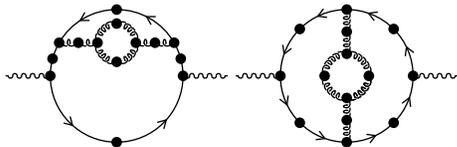}}}
  \caption{\footnotesize{A set of 3-loop diagrams that may give a dominant
  contribution, diagrams where the gluon loop is replaced by a fermion loop is
  also of the same nature.}}
  \label{fig:3looplog}
\end{figure}

Our result reveals also a logarithmic sensitivity, {\it i.e.} the gluon which is
supposed to be soft{\footnote{Where the use the effective propagator and the inherited
HTL approximation are justified}}, is extrapolated by the logarithmic behavior
($\int\limits_{gT}^{T}\frac{dl}{l}$) to the hard region, which bring again
the problem of the hard limit of the HTL approximation at next to leading order.
A complete result may need to include higher order contributions as the 3-loop
diagrams of Fig.~\ref{fig:3looplog}. Probably one has also to consider the other 
topologies of 3-loop diagrams to have a gauge independent set of diagrams, 
without forgetting the associated counterterms.
The importance multiple scattering of the quark in the plasma
(Landau-Pomeranchuk effect) in 
\cite{CleymGR1,CleymGR2,GolovR1}, points again the importance of higher order
diagrams. 

\section{Conclusion}
We have discussed the production of virtual soft dilepton production in a hot
quark-gluon plasma  in thermal equilibrium. This special case may give some
indicative properties of the hard thermal loop expansion as a 
perturbative scheme.
We have shown that to get the total contributions at a given order 
in the coupling constant $g$, one has to add contributions that comes from
different loop orders, {\it i.e.} increasing the number of loops does not necessarily
mean that the result is suppressed by additional powers of $g$. 
The case of soft virtual photon we considered illustrates that, even in the absence
of any divergences, loop-order mixing occurs. Moreover the two-loop diagrams give
rise to new physical processes which do not appear at one loop. Using
the method of counterterms we give an explicit way to calculate higher loop
diagrams using the HTL expansion. There exists other methods in the literature,
such as using a cutoff \cite{brayu,brathom,BaierNNR1,KapusLS1} in order to
obtain the correct behavior of the effective propagator in the hard momentum
limit and to avoid possible double counting.Finally adding the contributions 
coming from $L^{2}>0$ and $L^{2}<0$ the virtual photon production rate in the 
{\it plasma} is increased considerably which may have remarkable 
experimental effects.

\end{document}